\documentclass[oldversion]{aa}
\pdfoutput=1

\usepackage[intlimits]{amsmath}
\usepackage{txfonts}
\usepackage[american]{babel}
\usepackage{graphicx}
\usepackage{xspace}
\usepackage{upgreek}
\usepackage{multirow}

\usepackage{natbib}
\newcommand{\Alfven}{Alfv\'{e}n\xspace}
\newcommand{\simm}{\ensuremath{\mathord{\sim}}}
\newcommand{\Nabla}{\ensuremath{\vec{\nabla}}}
\newcommand{\abs}[1]{\ensuremath{\left|#1\right|}}
\newcommand{\vA}{\ensuremath{v_\mathrm{A}}}
\newcommand{\vAp}{\ensuremath{v_\mathrm{Ap}}}
\newcommand{\betain}{\ensuremath{\beta_\mathrm{i}}}
\newcommand{\cs}{\ensuremath{c_\mathrm{s}}}
\newcommand{\vc}{\ensuremath{v_\mathrm{c}}}
\newcommand{\cso}{\ensuremath{c_\mathrm{s0}}}
\newcommand{\eR}{\ensuremath{\hat{\vec{e}}_R}}
\newcommand{\er}{\ensuremath{\hat{\vec{e}}_r}}
\newcommand{\ex}{\ensuremath{\hat{\vec{e}}_x}}

\newcommand{\ez}{\ensuremath{\hat{\vec{e}}_z}}
\newcommand{\de}{\ensuremath{\mathrm{d}}}
\newcommand{\rc}{\ensuremath{r_\mathrm{c}}}
\newcommand{\tauK}{\ensuremath{\tau_\mathrm{K}}}
\newcommand{\tauM}{\ensuremath{\tau_\mathrm{M}}}
\newcommand{\const}{\ensuremath{\text{constant}}}
\newcommand{\OmegaO}{\ensuremath{\varOmega_0}}
\newcommand{\degree}{\ensuremath{^\circ}}

\newcommand{\pmag}{\ensuremath{p_\text{mag}}}
\newcommand{\mdotin}{\ensuremath{\dot{m}_\text{in}}}
\newcommand{\mean}[1]{\langle#1\rangle}
\newcommand{\mmean}[1]{\langle\!\langle#1\rangle\!\rangle}

\begin{document}

\title{Shearing box simulations of accretion disk winds}
\author{R.~Moll}
\date{Accepted on October 10, 2012}
%\offprints{}
\institute{Max-Planck-Institut f\"{u}r Astrophysik, Karl-Schwarzschild-Stra{\ss}e 1, 85748 Garching, Germany \and
Department of Astronomy and Astrophysics, University of California, Santa Cruz, CA 95064, USA}
\abstract{The launching process of a magnetically driven outflow from
an accretion disk is investigated in a local, shearing box model which allows a
study of the feedback between accretion and angular momentum loss. The
mass-flux instability found in previous linear analyses of this problem is
recovered in a series of 2D (axisymmetric) simulations in the MRI-stable (high
magnetic field strength) regime.  At low field strengths that are still
sufficient to suppress MRI, the instability develops on a short radial length
scale and saturates at a modest amplitude.  At high field strengths, a
long-wavelength ``clump'' instability of large amplitude is observed, with
growth times of a few orbits.  As speculated before, the unstable connection
between disk and outflow may be relevant for the time dependence observed in
jet-producing disks.  The success of the simulations is due in a large part to
the implementation of an effective wave-transmitting upper boundary
condition.}

\keywords{Magnetohydrodynamics (MHD) -- Accretion: accretion disks -- Instabilities}
\maketitle

\section{Introduction}

The occurrence of strong, collimated outflows in association with the accretion
of compact astronomical objects is a common phenomenon. Examples are jets from
young stellar objects, interacting binaries, active galactic nuclei and
possibly the central engines of gamma ray bursts.  Although they are often
regarded as separate entities, accretion disks and jets are physically
dependent on one another.  The mass loading of the jet depends on properties of
the disk and its immediate atmosphere, and the outflow feeds back on the disk
by carrying off material and angular momentum. The physics of the transition
from accretion flows to outflows is therefore a key element of our
understanding of both disks and jets.

An inflow of gas in a rotating disk necessitates the presence of a mechanism
for angular momentum transport.  It is well known that weak magnetic fields are
responsible for magnetorotational instabilities
\citep[MRI;][]{1991Balbus,2003Balbus}.  The turbulence driven by MRI leads to
internal stresses that redistribute angular momentum in a similar way as would
do a high viscosity, thus driving accretion.  MRI in disks have been the
subject of numerous numerical studies, including local simulations in shearing
boxes as well as global simulations
\citep[e.g.,][]{2002Hawley,2002Turner,2009Fromang}.

Accretion disks may function without MRI, or other forms of viscosity, if they
lose angular momentum through outflows. Such outflows are generated readily in
disks threaded by large-scale poloidal magnetic fields
\citep{1976Bisnovatyi,1976Blandford}.  The material is accelerated outwards
along the poloidal field by centrifugal forces. The inclination of the field
with respect to the disk determines the efficiency of the process. It must be
sufficiently small, $\mathord{<}60\degree$ in the cold wind model of
\citet{1982Blandford}, for the ``magnetic slingshot'' to work.  In an optically
thick disk with an isothermal atmosphere and a mean field that is sufficiently
strong to suppress MRI, the strongest winds are expected for inclinations of
$\mathord{\approx}45\degree$ and the mass loss rate decreases with increasing
field strength; in turbulent disks with weaker fields, the mass loss rate is
expected to increase with the field strength \citep{2001Ogilvie}.

Due to the way in which the inclination of the poloidal magnetic field is
coupled with the strength of the outflow, the connection between accretion and
magnetically driven outflows might be inherently unstable.  A possible
instability scenario is the following: an increase of the speed of radial
inflow increases the inclination of the field towards the midplane, which
increases the mass flux and the loss of angular momentum and thus leads to an
even faster inflow.  The existence of instabilities in disks that lose angular
momentum through a magnetically driven wind was first predicted by
\citet{1994Lubow}, disputed by \citet{1996Koenigl}, and confirmed by the
perturbation analyses of \citet{2002Cao} and \citet{2009Campbell}.

The perspective on the physics of the disk-wind connection taken here starts
with the stationary, one-dimensional flow problem of \citet{2001Ogilvie}, which
yielded the dependence of the outflow rate on the strength and inclination of
the magnetic field. A logical step towards a time-dependent view are linear
stability analyses of this stationary problem, such as the axisymmetric,
short-wavelength problem studied in \citet{2002Cao}.  The logical step taken
here is a two-dimensional, finite amplitude study of the same problem by
numerical simulations.

Borrowing from early studies of MRI, the problem is kept local in the radial
direction by using a Cartesian shearing box. The simulations can thus be seen
as an extension of 2D MRI simulations with a net magnetic flux threading the
box. The extension consists of including an outflow through the upper boundary,
and a finite asymptotic inclination of the magnetic field. In this way, it
provides a natural connection between MRI and wind-launching studies like those
of Ogilvie \& Livio.

The model is described in Sect.~\ref{sec:model}. Details of the numerical
realization are given in Sect.~\ref{sec:methods}.  The results of the
simulations are presented in Sect.~\ref{sec:results}, followed by a summary and
discussion in Sect.~\ref{sec:discussion}.

\section{The model}
\label{sec:model}

We consider an axisymmetric accretion disk threaded by a poloidal magnetic
field that is bent away from the rotation axis.  The accreting matter, which is
initially orbiting the central object in (near) radial force equilibrium, loses
angular momentum through a centrifugally accelerated wind, thus enabling an
accretion inflow.  We ignore the possibility of angular momentum transport by
viscosity or MRI, and, with one exception, assume no explicit magnetic
diffusivity.

The atmosphere of the disk is assumed to have a similar temperature as the disk
itself. Both density and plasma-$\beta$ decrease strongly with height.  Inside
the disk, $\beta$ is of order unity or larger, and the frozen-in magnetic field
is dragged along with the rotating plasma.  Above the disk, the magnetic field
is strong enough to enforce an approximately constant angular velocity along
the field lines.  The material trapped on the magnetic field is flung outwards
by centrifugal forces. Around the point where the flow reaches \Alfven speed,
the inertia of the gas dominates again and causes a strong winding of the
magnetic field into a predominantly azimuthal field.

The simulations are done in a Cartesian, periodic shearing box (see, e.g., the
MRI simulations by \citealt{1995Hawley}). The focus is thus on local processes,
happening on a radial scale of a few times the thickness of the disk and
smaller. The goal is to study the nonlinear development of short-wave processes
such as the instabilities predicted by linear analyses of the wind-launching
problem. A major advantage of the shearing box is that the calculations can be
done without adding an explicit magnetic diffusivity. With (quasi-)periodic
boundary conditions in the radial direction, magnetic flux does not pile up by
advection, as would happen in a global model. In effect, the shearing box model
thus elegantly incorporates a separation of time scales: it includes processes
that scale with the orbital period while leaving out long-term processes such
as the pile-up of magnetic flux.

Since a wind is to be launched, a realistic upper boundary that allows mass
outflow is needed. At low mass fluxes, the magnetic field configuration above
the disk is affected only little by the presence of the outflow. It is
therefore natural to approximate the poloidal field at the upper boundary as a
potential field.  The implementation of this boundary turns out to be critical
for the success of the model.

The mass loss from the box causes conditions to change slowly with time.  In
order to separate such secular changes from the processes under study, we add a
mass source close to the midplane, as in previous local models
\citep{2012Ogilvie}.  As an approximation, this should be valid under
conditions where the time scale for mass loss is long compared with the
accretion time scale.

\subsection{Magnetic support against gravity}
\label{sec:suppgrav}

The poloidal field in this model is not force-free.  Near the midplane, the
bending of the field generates a curvature force which counteracts gravity.
The midplane equilibrium is thus determined by inward gravity, outward
centrifugal force and outward magnetic curvature force.  Let \begin{equation}
\epsilon = \frac{B^2 / (4\pi\rho\rc) }{GM / R^2} = \frac{B^2 / (4\pi\rho\rc)
}{\varOmega^2 R} \end{equation} be an estimate for the relative importance of
the curvature force at the midplane if the rotation is Keplerian.  Taking the
curvature radius $\rc$ to be of the order of the disk's scale height $H$ and
estimating $\varOmega^2$ with \(\cs^2/H^2=p/H^2\rho\) (thin disk
approximation), we find \(\epsilon \sim \delta/\beta\).  For \(\beta\gtrsim 1\)
and a small aspect ratio \(\delta = H/R\), the curvature force has only minor
significance. It perturbs the radial equilibrium of a Keplerian disk only
slightly.  This difference is important, however, for the launching of the
magnetically powered wind \citep{1998Ogilvie}.

\section{Methods}
\label{sec:methods}

We use the shearing box approach \citep[e.g.,][]{1995Hawley} with axisymmetry
to solve the MHD equations in a Cartesian box that rotates with Keplerian
angular velocity \(\vec{\varOmega}_0=\OmegaO\ez\) at some distance $R_0$ from
the axis of rotation.  Compared to the dimensions of the box, $R_0$ is assumed
to be large enough that the equations to be solved become independent of it.

The majority of simulations was done assuming ideal MHD, for which the
induction equation has the usual form \(\partial_t\vec{B} = \Nabla \times
(\vec{v} \times \vec{B})\).  The momentum equation is
\begin{multline}
    \frac{\de \vec{v}}{\de t}
        = -\frac{\Nabla p}{\rho} + \frac{(\Nabla\times\vec{B})\times\vec{B}}{4\pi\rho} \\
        - 2 \vec{\varOmega}_0 \times\vec{v} + \OmegaO^2 ( 3 x \ex - z \ez )
        - \frac{3}{2} x \OmegaO M,
    \label{eq:mom}
\end{multline}
where \(x=R-R_0\) is the radial coordinate and $z$ is the height (vertical)
coordinate, $z=0$ corresponding to the midplane. The first two terms after the
Lorentz force are the result of a Taylor expansion of the centrifugal force \(R
\OmegaO^2 \eR\) and the gravitational acceleration \(-\OmegaO^2 R_0^3
r^{-2}\er\), assuming that \(\abs{x},\abs{z}\ll R_0\).  The last term accounts
for the momentum of material which is added through a mass source term
(described below). We also calculated diffusive cases in which the induction
and momentum equations are extended by $\eta\nabla^2\vec{B}$ and
$\nu\nabla^2\vec{v}$, respectively.

To prevent the box from being drained by the outflow, we introduce a mass
source term on the right-hand side of the continuity equation:
\begin{equation}
    \frac{\partial \rho}{\partial t} + \Nabla \cdot (\rho \vec{v}) = f \frac{\rho(t=0)}{\tauM} \eqqcolon M
    \label{eq:masssrc}
\end{equation}
where $f=1$ for $\abs{z} < 0.5$, $f=\cos[\pi(\abs{z}-0.5)]$ for $0.5 < \abs{z}
< 1$ and $f=0$ for $\abs{z} > 1$.  Integrated in $z$, the material introduced
in the time span $\tauM$ amounts to \(\varSigma_\mathrm{M}/\varSigma_0=61\%\)
of the surface density in the initial state. It is given the initial
temperature and Keplerian momentum (terms with $M$ in Eqs.~\ref{eq:mom}
and~\ref{eq:ene}). It damps horizontal and vertical motions and stabilizes the
temperature.

We adopt an ideal gas equation of state with \(\gamma=5/3\) and evolve the
equation for the internal energy density \(e = p / (\gamma-1)\),
\begin{equation}
    \frac{\partial e}{\partial t} + \Nabla \cdot (e\vec{v}) = - p \Nabla \cdot \vec{v} + K
    + M \frac{e(t=0)}{\rho(t=0)},
    \label{eq:ene}
\end{equation}
along with the corresponding equation for conservation of total energy.  In
places where the latter yields unphysical results (``negative pressures'' due
to discretization errors which occur occasionally in very low-$\beta$ regions),
we use the value obtained by evolving the internal energy instead of the total
energy.  The last term in Eq.~\eqref{eq:ene} accounts for the material which is
added through the mass source term. Since a calculation can become unfeasible
if regions with extremely low density develop, we add another artificial source
term
\begin{equation}
    K = \frac{T(t=0)-T(t)}{T(t=0)} \cdot \frac{e(t=0)}{\tauK}
\end{equation}
to the energy equation. This intervention helps to avoid low-density cavities
by relaxing the temperature to that of the initial state on an appropriate time
scale $\tauK$. In nature, radiative losses would prevent excessive
temperatures.

\subsection{Boundary conditions}

We assume reflective symmetry at the midplane. The computational domain is
\(-L_x < x < L_x\) in the horizontal direction and \(0 < z < L_z\) in the
vertical direction. The bottom boundary corresponds to the accretion disk's
midplane.  There, we use reflective boundary conditions: \(\partial
(\rho,T,v_{[x,y]},|v_z|,|B_{[x,y]}|) / \partial z = 0\), the signs of $v_z$ and
$B_{[x,y]}$ are reversed and $B_z$ follows from solenoidality.

The horizontal boundaries are strictly periodic for all quantities except the
azimuthal velocity, for which
\begin{equation}
    v_y(x,z) = v_y(x \pm 2 L_x,z) \pm \frac{3}{2} L_x\OmegaO
\end{equation}
is applied in the ghost cells of the left (right) boundary.

\subsubsection{Top boundaries}
\label{topboundaries}

The top boundary conditions ($z>L_z$) must account for the effects of a global
magnetic field outside the scope of the local simulation and allow for an
unhindered outflow of material.  We tried different ways of implementing these
constraints and found many to be unfit: zero-slope conditions in the vertical
direction create numerical instabilities and rigorously imposing the poloidal
field inclination angle causes significant reflections, in some cases strongly
affecting the dynamics in the simulated domain.

The conditions described below are constructed in view of the limiting case far
above the midplane. There, $\beta\ll 1$, which suggests that the magnetic field
is force-free, and the characteristics of MHD waves along the field are
outwards, which suggests the use of extrapolation along the magnetic field.  A
convenient assumption then is that the poloidal magnetic field is a potential
one\footnote{It should be borne in mind, though, that this may not be a good
approximation in cases where the azimuthal field exerts strong Lorentz forces
perpendicular to the poloidal field.}.  An inclination angle is imposed by
choosing the value of the mean field. This gives the field more freedom than a
strict imposition of the inclination angle. These boundary conditions work very
well even in cases where the boundary is not far in the $\beta\ll 1$ domain,
still inside the \Alfven surface.  They are numerically stable and cause only
minimal reflective artifacts.

The conditions are implemented as follows.  At each time step, we determine the
potential field that matches with $B_z$ at $z=L_z$ and satisfies
$[\Nabla\times\vec{B}]_y=0$.  The solution is found by means of a discrete
Fourier transform in $x$-direction, using $\hat{B}_x(k) = - i \hat{B}_z(k)$ for
the complex Fourier coefficients of $B_x$ ($\hat{B}_z$ being the coefficients
of $B_z$) for $k\ne0$.  The mean field (corresponding to $k=0$) is chosen such
that at infinity the poloidal field is inclined by an angle $\iota$ with
respect to the horizontal.  We use zero-slope conditions along the magnetic
field: with $s$ measuring distance along a poloidal field line, $\partial
(\rho,v_{[x,y,z]},B_y)/\partial s = 0$ to first order (i.e., the values are
constant along a field line).  In addition, we dismiss inertial and
gravitational forces parallel to the magnetic field in the uppermost layer of
cells next to the upper boundary.  The temperature is kept fixed at the top
boundary, $\partial T/\partial t = 0$.

\subsection{Initial conditions}
\label{sec:inicond}

The gas is initially isothermal, with both pressure and density being
\(\mathord{\propto} \exp[-z^2/(2l_0^2)]\), where $l_0$ is the unit length
(i.e., $l_0$ is the scale height in the initial state; we shall denote the
\emph{actual}, time dependent scale height of the disk with $H$).  The initial
magnetic field is homogeneous and inclined by an angle \(\iota\) with respect
to the midplane.  Its strength is determined by the simulation parameter
$\betain$, which represents the ratio of the gas-to-magnetic pressure at the
midplane in the initial state.  The initial velocity is
Keplerian\footnote{Taking magnetic support against gravity
(Sect.~\ref{sec:suppgrav}) into account would change $v_y$ by a constant amount
\(-\epsilon R_0 \OmegaO /2\). We opted not to include such a correction and let
the system find an equilibrium by itself instead.} at the midplane, \(v_y =
-3/2 x \OmegaO\), and constant along the magnetic field lines.

\subsection{MHD code}

The simulations were performed with the Eulerian MHD code of
\citet{2008Obergaulinger}.  It is based on a flux-conservative finite-volume
formulation of the MHD equations and a constraint transport scheme that
maintains a divergence-free magnetic field \citep{1988Evans}.  Using
high-resolution shock capturing methods \citep[e.g.,][]{1992LeVeque}, it allows
a choice of various optional high-order reconstruction algorithms and
approximate Riemann solvers based on the multi-stage method \citep{2006Toro}.
The simulations presented here were performed using a fifth order
monotonicity-preserving reconstruction scheme \citep{1997Suresh}, the HLL
Riemann solver \citep{1983Harten}, and third order Runge-Kutta time stepping.

\section{Results}
\label{sec:results}

\begin{table*}
\caption{List of simulations and mean properties in the evolved state}
\centering
\begin{tabular}{c|cc|ccccccc}
\hline\hline
\multirow{2}{*}{Run\tablefootmark{a}} &
Domain&
\multirow{2}{*}{Resolution}&
\multirow{2}{*}{$\beta=\frac{\mmean{p}}{\mmean{\pmag}}$} &
$\mmean{\varSigma}$ &
$\mmean{H}$ &
$\mmean{\rho v_z}$ &
$\mmean{\rho v_z\Delta v_y}$ &
$\mmean{-B_zB_y/4\pi}$ &
$\mmean{\mdotin}$
\\
&
$[l_0]$&
&
&
$[\varSigma_0]$ &
$[l_0]$ &
$[\mmean{\varSigma}\OmegaO]$ &
$[\mmean{\Sigma}\mmean{H}\OmegaO^2]$ &
$[\mmean{\Sigma}\mmean{H}\OmegaO^2]$ &
$[m_{\mmean{H}}\OmegaO]$
\\
\hline
  $\beta1\,\iota45$ & $4\times2$ & $256\times128$ & $2.76_{\pm8\%}$ & $0.718_{\pm3\%}$ & $0.651_{\pm3\%}$
        & $0.0677_{\pm29\%}$ & $0.0897_{\pm65\%}$ & $1.41_{\pm13\%}$ & $3.07_{\pm17\%}$ \\ % D1 (20 orbits)
  $\beta1\,\iota45\,$h & $4\times2$ & $512\times256$ & $2.43_{\pm8\%}$ & $0.642_{\pm2\%}$ & $0.668_{\pm6\%}$
        & $0.0756_{\pm28\%}$ & $0.108_{\pm82\%}$ & $1.15_{\pm25\%}$ & $2.56_{\pm60\%}$ \\ % C1v7_hires (7)
  $\beta1\,\iota45\,$h$\nu$ & $4\times2$ & $512\times256$ & $2.55_{\pm4\%}$ & $0.695_{\pm4\%}$ & $0.668_{\pm2\%}$
        & $0.0687_{\pm8\%}$ & $0.0807_{\pm10\%}$ & $1.27_{\pm7\%}$ & $2.71_{\pm6\%}$ \\ % D1v7b (7)
  $\beta1\,\iota45\,$h$\eta$ & $4\times2$ & $512\times256$ & $3.19_{\pm6\%}$ & $0.789_{\pm1\%}$ & $0.639_{\pm4\%}$
        & $0.0615_{\pm20\%}$ & $0.0865_{\pm76\%}$ & $1.61_{\pm8\%}$ & $3.48_{\pm19\%}$ \\ % D1v7c_eta (8)
  $\beta1\,\iota45\,$b & $6\times3$ & $384\times192$ &  $2.56_{\pm20\%}$ & $0.624_{\pm1\%}$ & $0.661_{\pm15\%}$
        & $0.0744_{\pm29\%}$ & $0.111_{\pm72\%}$ & $1.31_{\pm27\%}$ & $2.97_{\pm34\%}$ \\ % D1v5_bigger (7)
  $\beta1\,\iota37$ & $4\times2$ & $256\times128$ & $3.76_{\pm18\%}$ & $0.586_{\pm3\%}$ & $0.557_{\pm13\%}$
        & $0.0831_{\pm39\%}$ & $0.177_{\pm55\%}$ & $1.93_{\pm20\%}$ &$ 4.27_{\pm40\%}$ \\ % D2b (10)
  $\beta1\,\iota40$ & $4\times2$ & $256\times128$ & $3.27_{\pm16\%}$ & $0.626_{\pm2\%}$ & $0.583_{\pm9\%}$
        & $0.0775_{\pm23\%}$ & $0.143_{\pm42\%}$ & $1.70_{\pm21\%}$ & $3.79_{\pm27\%}$ \\ % C2 (7)
  $\beta1\,\iota50$ & $4\times2$ & $256\times128$ & $1.54_{\pm14\%}$ & $0.762_{\pm10\%}$ & $0.723_{\pm9\%}$
        & $0.0677_{\pm83\%}$ & $0.118_{\pm174\%}$ & $0.617_{\pm30\%}$ & $1.51_{\pm59\%}$ \\ % C3 (5)
  $\beta1\,\iota60$ & $4\times2$ & $256\times128$ & $1.89_{\pm14\%}$ & $1.15_{\pm11\%}$ & $0.944_{\pm5\%}$
        & $0.0428_{\pm60\%}$ & $0.0456_{\pm110\%}$ & $0.190_{\pm59\%}$ & $0.494_{\pm183\%}$ \\ % C3b (10)
$\beta0.5\,\iota45$ & $4\times2$ & $256\times128$ & $1.69_{\pm35\%}$ & $0.909_{\pm16\%}$ & $0.627_{\pm20\%}$
        &  $0.0529_{\pm135\%}$ & $0.0951_{\pm317\%}$ & $1.21_{\pm104\%}$ & $2.64_{\pm140\%}$ \\ % D4 (15)
  $\beta2\,\iota45$ & $4\times2$ & $256\times128$ &$ 3.46_{\pm6\%}$ & $0.486_{\pm2\%}$ & $0.698_{\pm3\%}$
        &  $0.100_{\pm15\%}$ & $0.147_{\pm31\%}$ & $1.05_{\pm11\%}$ & $2.46_{\pm15\%}$ \\  % C5 (10)
  $\beta8\,\iota45$ & $4\times2$ & $256\times128$ & $7.12_{\pm1\%}$ & $0.286_{\pm0\%}$ & $0.793_{\pm1\%}$
        &  $0.170_{\pm2\%}$ & $0.219_{\pm4\%}$ & $0.447_{\pm2\%}$ & $1.34_{\pm1\%}$ \\  % C5b (10)
 $\beta10\,\iota45$ & $4\times2$ & $256\times128$ & $7.66_{\pm9\%}$ & $0.270_{\pm2\%}$ & $0.806_{\pm3\%}$
        &  $0.180_{\pm8\%}$ & $0.227_{\pm13\%}$ & $0.480_{\pm7\%}$ & $1.42_{\pm8\%}$ \\ % C5bb (10)
\hline
\end{tabular}
\tablefoot{
The quantities in columns 4--10 are average values in the evolved state (for a
visualization, see Fig.~\ref{fig:sims}).  Double angle brackets
$\mmean{\ldots}$ denote a combined temporal and horizontal mean.  Temporal
fluctuations are rounded to full percents.  \tablefoottext{a}{The label
contains the initial value of $p/\pmag$ at the midplane (denoted $\betain$ in
the text) and the asymptotic inclination angle $\iota$ of the potential field
at the outer vertical boundary ($0\degree$ is horizontal and $90\degree$ is
vertical).  In addition, ``h'' stands for high resolution, ``$\nu$'' stands for
the use of an explicit shear viscosity, ``$\eta$'' stands for the use of
explicit magnetic diffusion and ``b'' stands for a bigger computational box.}
}
\label{tab:sims}
\end{table*}

\begin{figure}[t]
\centering
\includegraphics[width=.9\linewidth]{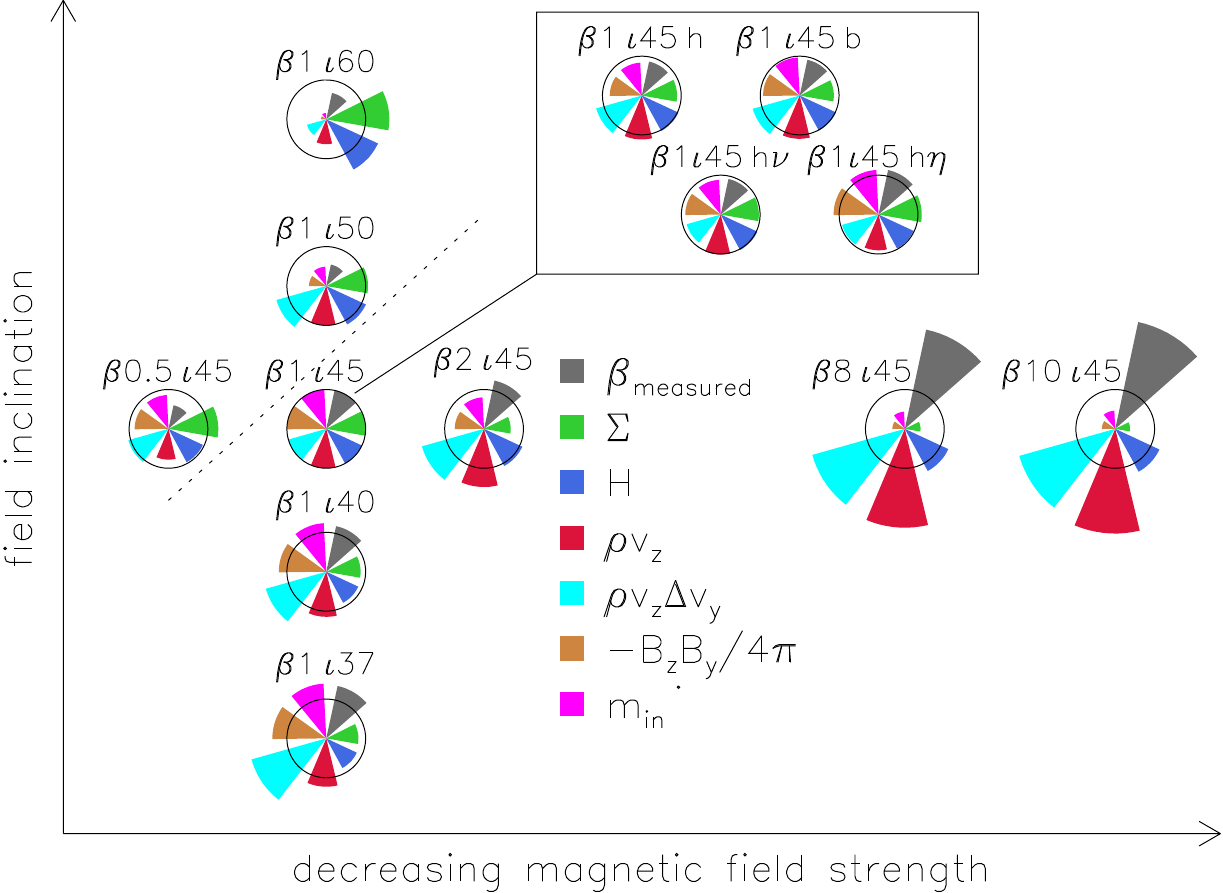}
\caption{Visualization of the wind properties listed in Table~\ref{tab:sims}.
Each ``flower'' represents a simulation.  The radii of the ``petals'' are
proportional to the represented value.  All values are normalized by the
respective value in run $\beta1\iota45$.  Strong clumps develop in the cases to
the left and above the dotted line (cf. Figs.~\ref{fig:C1map}, \ref{fig:Cimaps}
and~\ref{fig:Cbmaps}). Note that the $\beta$ in the simulation identifiers
refers to the initial state and measures the absolute field strength, whereas
the $\beta$ represented by the gray petals refers to the actual value in the
evolved state (see Table~\ref{tab:sims} and text).}
\label{fig:sims}
\end{figure}

\begin{figure}[t]
\centering
\includegraphics[width=.8\linewidth]{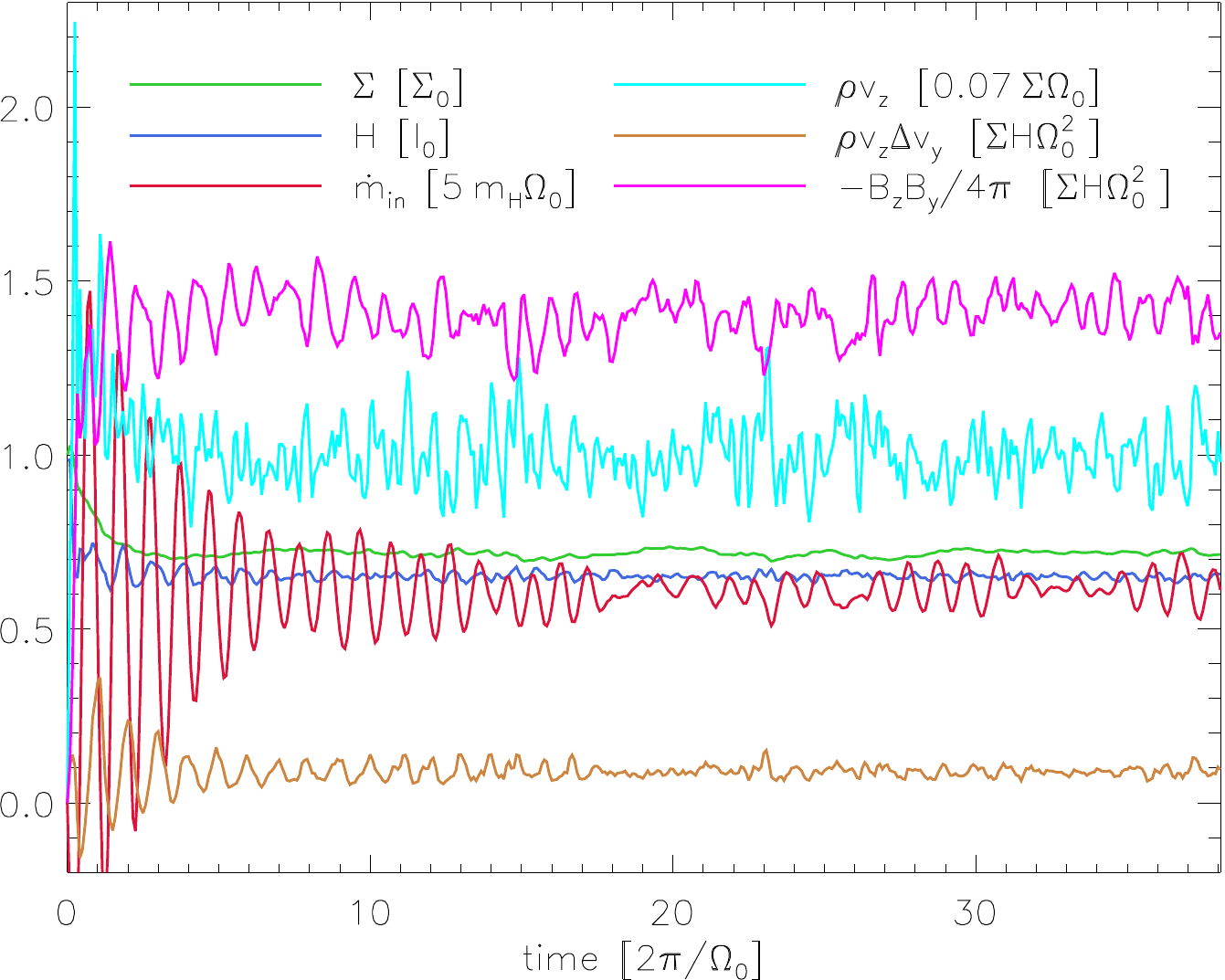}
\caption{Horizontal means of the surface density (green line), scale height
(blue line), mass inflow rate (red line), vertical mass flux (cyan line),
Reynolds stress (brown line) and Maxwell stress (purple line) as functions of
time in run $\beta1\iota45$.}
\label{fig:C1summ}
\end{figure}

\begin{figure}[t]
\centering
\includegraphics[width=.8\linewidth]{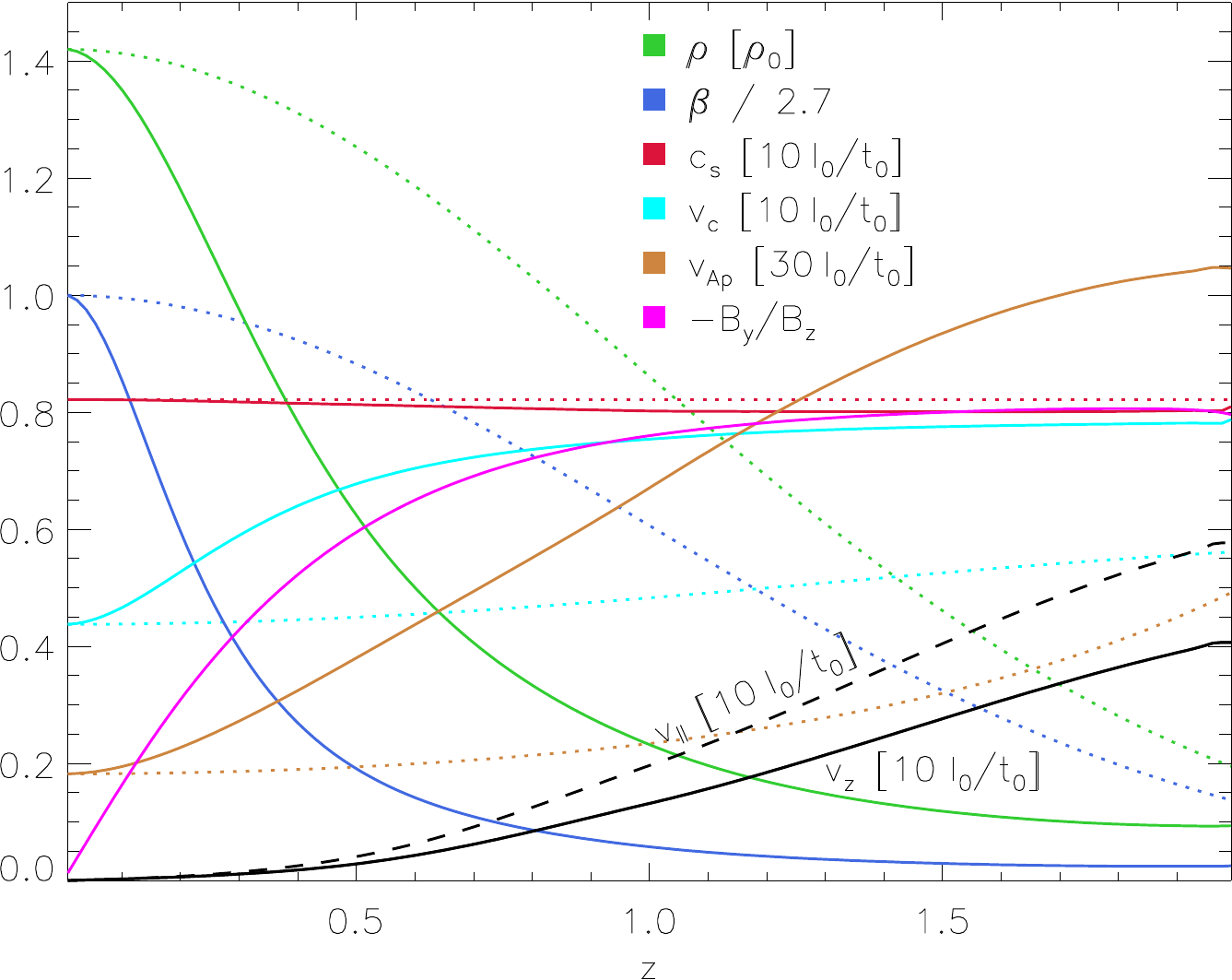}
\caption{Vertical stratification of the disk in run $\beta1\iota45$: density
(green line), plasma-$\beta$ (blue line), sound speed ($\propto$ square root of
temperature, red line), slow magnetosonic cusp speed (cyan line), poloidal
\Alfven velocity (brown line) and azimuthal magnetic field (magenta line). The
solid and dashed black lines represent the vertical velocity and the velocity
along the poloidal magnetic field, respectively.  The curves represent the mean
over the horizontal direction and time in the evolved state. For comparison,
dotted lines represent the isothermal, Gaussian stratification of the initial
condition (in arbitrary normalization).}
\label{fig:C1prof}
\end{figure}

\begin{figure}[t]
\centering
\includegraphics[width=.9\linewidth]{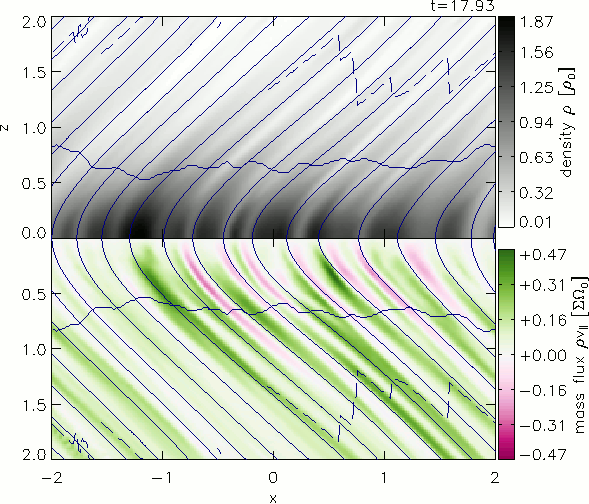}
\caption{Snapshot of the density (black/white) and the field-parallel mass
flux (green/pink colors) in run $\beta1\iota45$. Selected magnetic field lines
are drawn in blue.  The additional solid and dashed lines indicate the scale
height and the sonic surface, respectively.}
\label{fig:C1map}
\end{figure}

We performed several simulations with varying physical and numerical
parameters.  Long-lived runs, in which the disk evolution could be followed
over many orbits past an initial transient, are listed in Table~\ref{tab:sims}.
The time scale $\tauK$ for the temperature control was always taken to be $0.1$
orbits, so that large deviations from the initial isothermal state are rare;
e.g., in run $\beta1\iota45$, only about 4\% of the volume deviates from the
initial temperature by more than 10\%.  Unless noted otherwise, the time scale
$\tauM$ for the mass source used was $2$ orbits ($=2\pi/\OmegaO \equiv t_0$,
used as unit time).  In a simulation similar to $\beta1\iota45$ but without a
mass source ($\tauM\rightarrow\infty$), the mass inside the box is diminished
by $58\%$ during 2 orbits (after which the increased \Alfven velocities make it
impractical to continue the simulation).  The material introduced with the
$\tauM=2$ mass source is roughly of this value (cf.
$\varSigma_\mathrm{M}/\varSigma_0$ in Sect.~\ref{sec:methods}).  The
simulations were continued to $\simm20$ orbits, at which time the system has
passed through the initial transient and has entered a (more or less, depending
on the parameters) quasi-stationary state.

The two basic simulation parameters are $\betain$, which determines the
strength of the magnetic field, and $\iota$, which determines the asymptotic
inclination of the poloidal magnetic field in the disk's magnetosphere.
$\betain$ is usually smaller than the mean value of $\beta$ at the midplane in
the evolved state, which is listed in the 4th column of Table~\ref{tab:sims}.
An exception are disks with very weak magnetic fields, see
Sect.~\ref{sec:weakfields}.

Fig.~\ref{fig:C1summ} shows the evolution of horizontally averaged quantities
in run $\beta1\iota45$.  In general, the simulations start with strong
epicyclic oscillations in the inflow rate, presumably governed by the interplay
of magnetic curvature forces (due to the bending of the magnetic field at the
midplane, see Sect.~\ref{sec:suppgrav}), the Coriolis force and the loss of
angular momentum by the magnetically powered wind.  The disk gradually enters a
new equilibrium state with a vertical outflow.  For further analyses, we
consider only the times after the initial transient.  Depending on the
simulation, the evolved state is more or less relaxed.  In some cases, the
oscillations are still strong after many orbits (see fluctuations given in
percent in Table~\ref{tab:sims}), with no decreasing trend at the end of the
simulated time span.

The green curve in Fig.~\ref{fig:C1summ} shows the surface density
$\varSigma=\int\rho\,\de z$, normalized by its initial value.  We consider it a
measure for the efficiency of the wind, with strong winds yielding a smaller
$\varSigma$.  In the evolved state, the mass loss through the wind is balanced
by the mass source, so that $\varSigma\approx\const$.  The blue curve shows the
scale height $H$, which we define here as the height that contains 68\% of the
mass (we take into account the finite vertical size of the computational box,
so that $H \equiv l_0$ in the initial state).  The cyan curve shows the
vertical mass flux near the upper boundary, measured at height $z=1.7$.  In the
evolved state, its mean value equals the amount of material that is introduced
by the mass source: $\rho v_z \approx \varSigma_\mathrm{M}/\tauM = 0.0676 \,
\varSigma \OmegaO$.  The brown and purple lines show the Reynolds stress $\rho
v_z\Delta v_y$, where $\Delta v_y$ denotes the deviation from the Keplerian
rotation velocity, and the Maxwell stress $-B_zB_y/4\pi$, respectively.  The
red curve shows the mass inflow rate $\mdotin = \int \rho v_x \, \de z$,
normalized such that it gives, in units of the scale height, the horizontal
width of the disk material that crosses a surface of constant radius per radian
rotation.

Fig.~\ref{fig:C1prof} shows (horizontally averaged) vertical profiles of
various quantities in run $\beta1\iota45$.  Density and plasma-$\beta$,
represented by the green and the blue line, decrease by factors of 15 and 50,
respectively, from the midplane to the upper boundary. Both quantities drop
considerably faster with height than in a Gaussian stratification.  The mean of
the sound speed, represented by the red line, is close to the initial value
(which is $\sqrt{\gamma} 2\pi l_0/t_0$) at every height.  Above the disk, it
approximates the slow magnetosonic cusp speed $\vc$, given by $\vc^2 = \cs^2
\vA^2 / (\cs^2+\vA^2)$ and represented by the cyan line.  The \Alfven velocity
of the poloidal field, $\vAp$, is represented by the brown line; it is by a
factor of four larger than the cusp speed at the upper boundary.  The magenta
line represents the azimuthal magnetic field component, normalized by the
vertical field (the mean of which is a constant).

Fig.~\ref{fig:C1map} shows snapshots of the density $\rho$ (in units of the
initial value at the midplane) and of the mass flux $\rho v_\parallel$ in the
direction of the poloidal magnetic field ($v_\parallel$ is computed from $v_z$
and the local field inclination) in run $\beta1\iota45$.  The outflow is
inhomogeneous, the mass flux varies across different field lines.  Dark-green
stripes indicate the regions where the wind is strongest. Along some of the
field lines, there is an inflow at low heights (pink colors).  As angular
momentum is lost through the wind, an accretion flow drags the magnetic field
inward (i.e., in $-x$-direction).  A time animation shows that the stripes move
inward with the magnetic field.  On a given field line, the strength of the
wind varies periodically with a frequency of $\simm1.2$ per orbit.

\subsection{Dependence on field strength and inclination}

\begin{figure*}[t]
\centering
\begin{tabular}{cc}
\parbox{.45\linewidth}{
(a) $\beta1\,\iota37$ \\
\includegraphics[width=.9\linewidth]{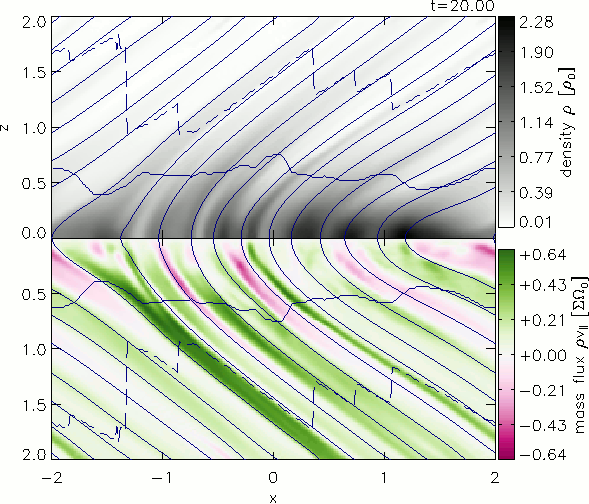} } &
\parbox{.45\linewidth}{
(b) $\beta1\,\iota40$ \\
\includegraphics[width=.9\linewidth]{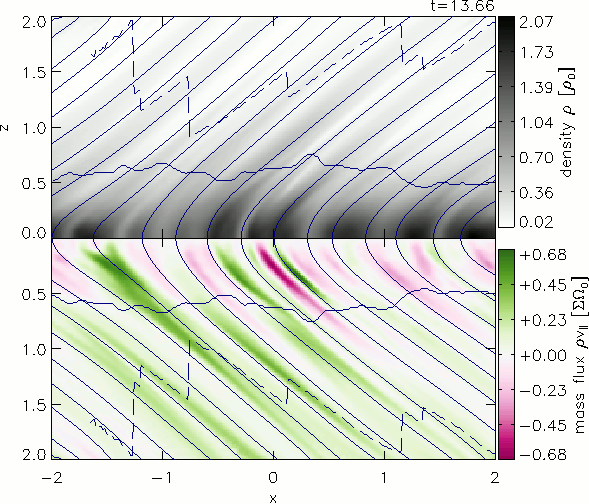} } \\
\parbox{.45\linewidth}{
(c) $\beta1\,\iota50$ \\
\includegraphics[width=.9\linewidth]{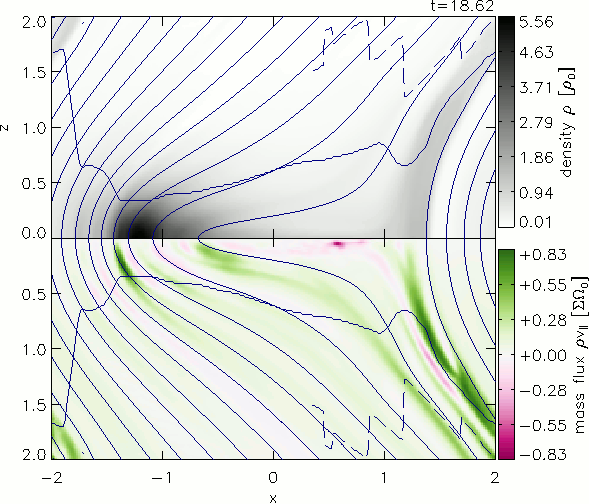} } &
\parbox{.45\linewidth}{
(d) $\beta1\,\iota60$ \\
\includegraphics[width=.9\linewidth]{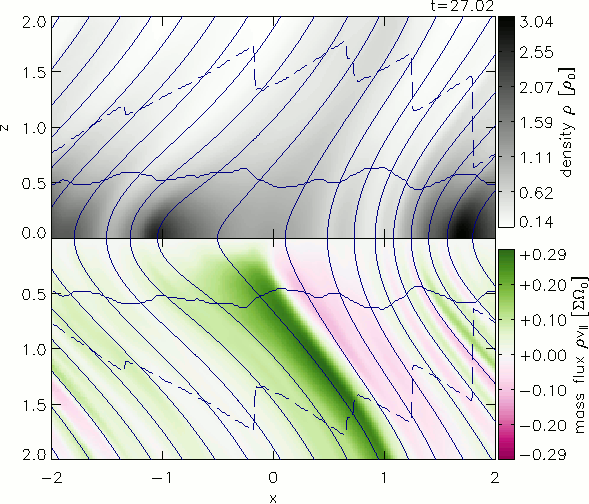} }
\end{tabular}
\caption{Snapshots of the density (black/white) and the field-parallel mass
flux (green/pink colors) in the evolved state of simulations with different
values for the asymptotic field inclination: (a) $\iota=37\degree$, (b)
$\iota=40\degree$, (c) $\iota=50\degree$, (d) $\iota=60\degree$.  The
$\iota=45\degree$ case is shown in Fig.~\ref{fig:C1map}. The plots include
selected magnetic field lines, the scale height (solid horizontal lines) and
the sonic surface (dashed lines).}
\label{fig:Cimaps}
\end{figure*}

\begin{figure*}[t]
\centering
\begin{tabular}{cc}
\parbox{.45\linewidth}{
(a) $\beta2\,\iota45$ \\
\includegraphics[width=.9\linewidth]{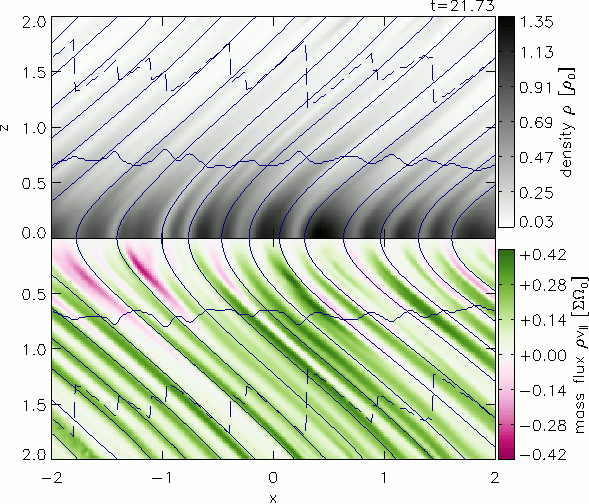} } &
\parbox{.45\linewidth}{
(b) $\beta0.5\,\iota45$ \\
\includegraphics[width=.9\linewidth]{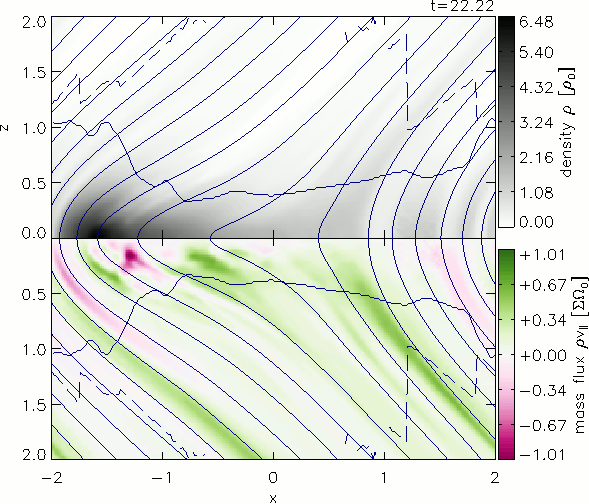} } \\
\parbox{.45\linewidth}{
(c) $\beta8\,\iota45$ \\
\includegraphics[width=.9\linewidth]{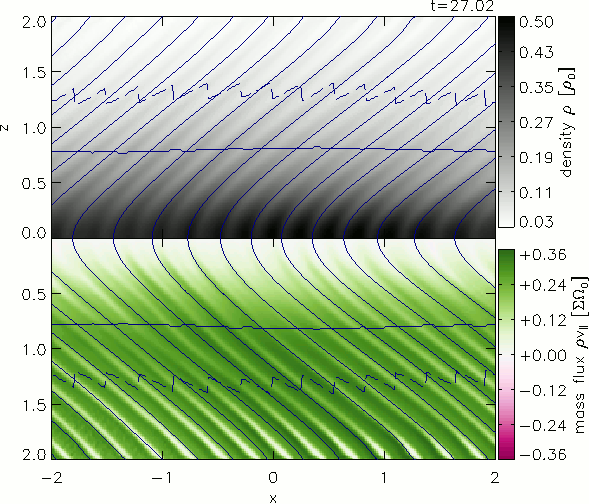} } &
\parbox{.45\linewidth}{
(d) $\beta10\,\iota45$ \\
\includegraphics[width=.9\linewidth]{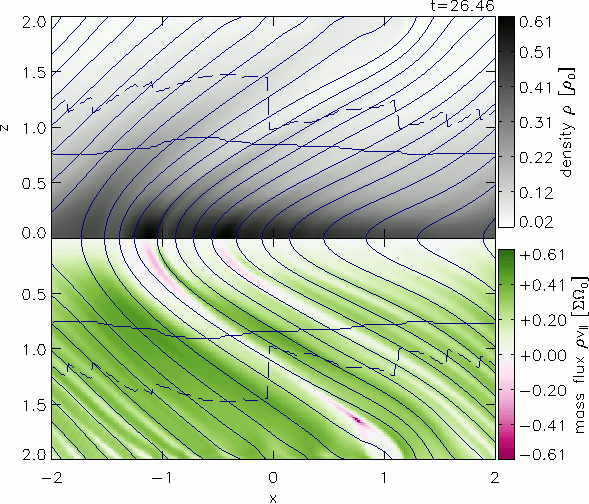} } \\
\end{tabular}
\caption{Snapshots of the density (black/white) and the field-parallel mass
flux (green/pink colors) in the evolved state of simulations with different
values for the initial gas-to-magnetic pressure: (a) $\betain=0.5$, (b)
$\betain=2$, (c) $\betain=8$, (d) $\betain=10$. The $\betain=1$ case is shown
in Fig.~\ref{fig:C1map}.  The plots include selected magnetic field lines,
the scale height (solid horizontal lines) and the sonic surface (dashed
lines).}
\label{fig:Cbmaps}
\end{figure*}

The temporal means of the quantities discussed above are listed in
Table~\ref{tab:sims} for different simulations.  A graphical representation of
these numbers can be found in Fig.~\ref{fig:sims}.  Snapshots of simulations
with different values for the asymptotic poloidal field inclination $\iota$ are
collected in Fig.~\ref{fig:Cimaps}, those with different values of $\betain$
are collected in Fig.~\ref{fig:Cbmaps}.

With increasing field inclination $\iota$, the strength of the wind, as
measured by the value of the surface density $\varSigma$ or the
density-normalized mass flux $\rho v_z$, decreases.  The disk is less compact
(increasing $H$) and the inflow rate smaller.  The disks in runs with high
inclination, $\iota=50\degree$ and $\iota=60\degree$, develop a large-scale
instability in the form of a density clump. We discuss the clump in the
$\iota=50\degree$ case in Sect.~\ref{sec:clumps}.

With increasing $\betain$ (decreasing field strength), $\varSigma$ decreases,
hence the wind becomes more effective at removing disk material. The inflow
rate $\mdotin$, however, is smaller.  While the Reynolds stress increases, the
inflow rate and the Maxwell stress, which is the dominant agent for the removal
of angular momentum, both decrease.  The wind is more homogeneous across the
magnetic field in the high-$\beta$ cases.

The mean value of $\beta$ at the midplane is usually higher than the initial
value $\betain$ as a result of the arbitrary initial transient.  For equal
asymptotic inclinations $\iota$, the field inclination as a function of height
is different in simulations with different field strengths. For instance, in
run $\beta8\iota45$, the horizontal separation between the foot point of a
field line at the midplane and the intersection point of the same field line at
the upper boundary is typically larger than in run $\beta1\iota45$, see
Figs.~\ref{fig:Cimaps}a and~\ref{fig:Cbmaps}c.  That is, the mean inclination
of the field is higher in run $\beta8\iota45$. In general, the inclination is a
function of the field line as well as height (or distance along the field
line).

\subsection{Resolution, diffusion and box size}

\begin{figure*}[t]
\centering
\begin{tabular}{cc}
\parbox{.45\linewidth}{
(a) $\beta1\,\iota45$\,b \\
\includegraphics[width=.9\linewidth]{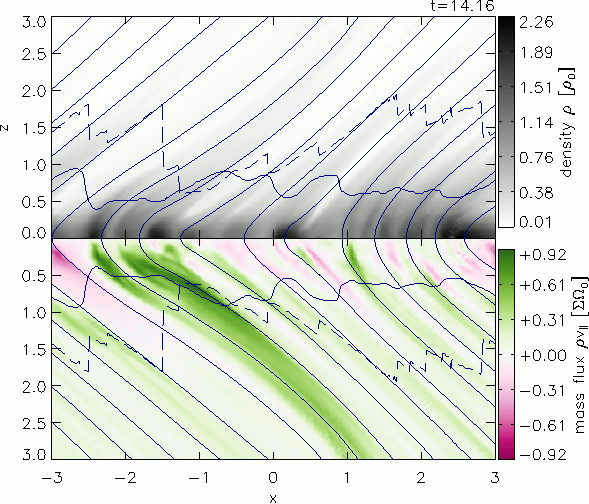} } &
\parbox{.45\linewidth}{
(b) $\beta1\,\iota45$\,h \\
\includegraphics[width=.9\linewidth]{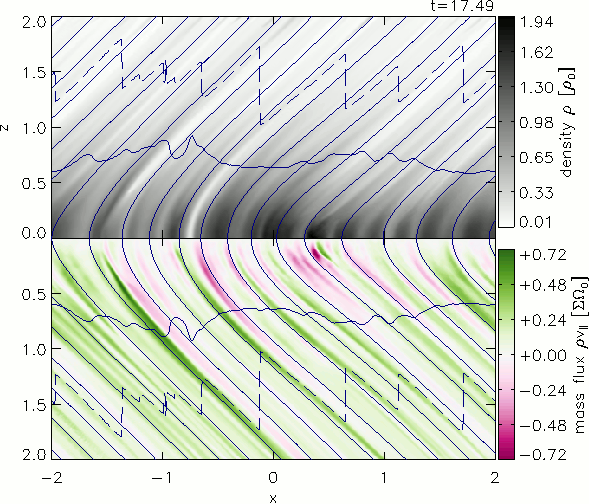} } \\
\parbox{.45\linewidth}{
(c) $\beta1\,\iota45$\,h$\nu$ \\
\includegraphics[width=.9\linewidth]{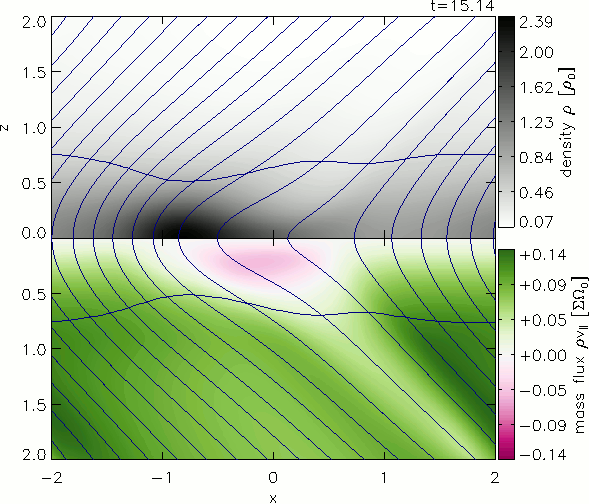} } &
\parbox{.45\linewidth}{
(d) $\beta1\,\iota45$\,h$\eta$ \\
\includegraphics[width=.9\linewidth]{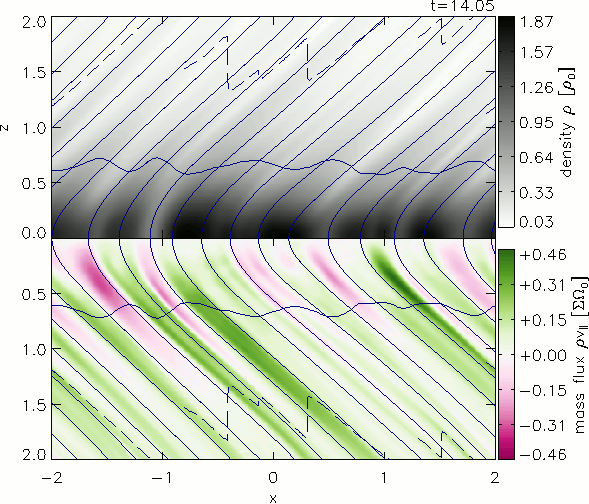} }
\end{tabular}
\caption{Snapshots of the density (black/white) and the field-parallel mass
flux (green/pink colors) in the evolved state of variants of run
$\beta1\iota45$ shown in Fig.~\ref{fig:C1map}. The panels show the effect of
(a) a bigger computational box, (b) a doubled resolution, (c) a doubled
resolution and explicit shear viscosity, (d) a doubled resolution and explicit
magnetic diffusivity.  The plots include selected magnetic field lines, the
scale height (solid horizontal lines) and the sonic surface (dashed lines).}
\label{fig:C1vmaps}
\end{figure*}

A comparison of simulations $\beta1\iota45$ and $\beta1\iota45$h,
Figs.~\ref{fig:C1map} and~\ref{fig:C1vmaps}b, shows that the inhomogeneities
(stripes) in the mass flux are narrower at higher resolution. The short
wavelengths appear to grow fastest, with numerical resolution the limiting
factor.

We study the effects of diffusion of magnetic field and momentum through a
high-resolution simulation ($\beta1\iota45$h$\eta$) with an explicit magnetic
diffusivity $\eta=10^{-3}\OmegaO l_0^2$ ($=10^{-3}\,\cso^2/\OmegaO$ with
$\cso=(p_0/\rho_0)^{1/2}$ being the isothermal sound speed in the initial
state) and another high-resolution simulation ($\beta1\iota45$h$\nu$) with an
explicit shear viscosity $\nu=10^{-2}\OmegaO l_0^2$.  In the case with magnetic
diffusivity, Fig.~\ref{fig:C1vmaps}d ($\beta1\iota45$h$\eta$), the stripes are
broader than in the low-resolution simulation without diffusivity.  In the case
with viscosity, Fig.~\ref{fig:C1vmaps}c ($\beta1\iota45$h$\nu$), the wind
inhomogeneities are mostly gone but the disk develops a clump.  The disk is
also more affected by instabilities in a simulation with a bigger domain
($\beta1\iota45$b), see Fig.~\ref{fig:C1vmaps}a.  The mean wind properties,
however, are similar in all these cases, see Fig.~\ref{fig:sims}. We conclude
that the amplitude of the instability increases with length scale, its shortest
length scale is limited by (real or numerical) diffusion, its largest length by
the box size.

\subsection{Replenishment of mass}
\label{sec:vartaum}

\begin{table}
\caption{Average value of the surface density $\varSigma$ $[\varSigma_0]$ in
runs with different $\tauM$}
\centering
\begin{tabular}{c|ccc}
\hline\hline
$\tauM$ & $(\betain,\iota)=(1,50\degree)$ & $(\betain,\iota)=(1,45\degree)$ & $(\betain,\iota)=(8,45\degree)$ \\
\hline
0.1 & 5.32 & 5.18 & 3.91 \\
2 & 0.762 & 0.718 & 0.286 \\
4 & 0.524 & 0.481 & 0.178 \\
10 & 0.376 & 0.259 & 0.106 \\
\hline
\end{tabular}
\label{tab:vartaum}
\end{table}

\begin{figure}[t]
\centering
\includegraphics[width=.75\linewidth]{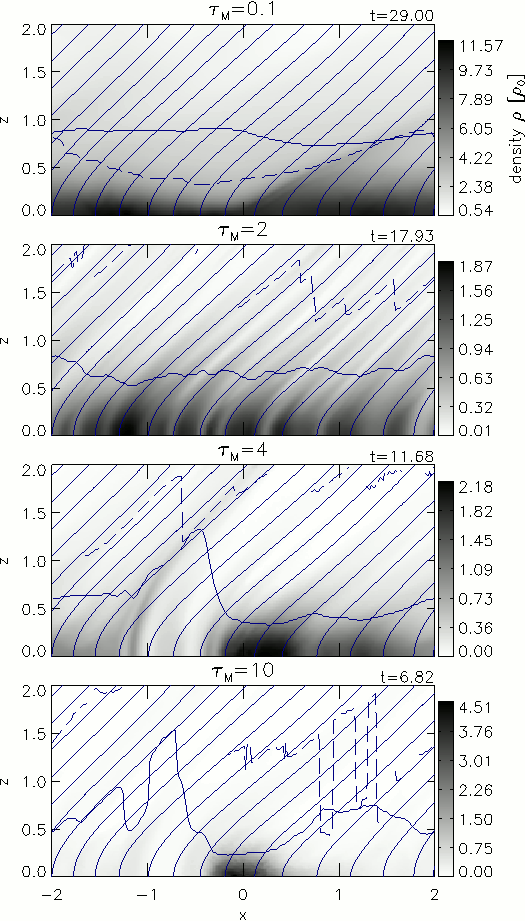}
\caption{Snapshots of the density in runs with $\betain=1$ and
$\iota=45\degree$ for different values of $\tauM$. The second panel from the
top shows the standard case ($\beta1\iota45$, compare Fig.~\ref{fig:C1map}).
The solid and dashed lines indicate the scale height and the sonic surface,
respectively.}
\label{fig:vartaum}
\end{figure}

In the simulations presented above, the value of $\tauM$ is always 2.  This not
only facilitates the comparison, but it also makes practical sense because it
keeps the surface density at values which are not very far from the initial
state (which is that of a classical thin disk).  Since the mass source is
somewhat artificial, it is reasonable not to have it too strong, where
``strong'' means putting in more mass than what would be lost without it in the
course of a few orbits.  On the other hand, simulations with very low density
are unfeasible for numerical reasons.

Table~\ref{tab:vartaum} shows how the surface density changes for different
values of the $\tauM$.  As expected, it increases with decreasing $\tauM$.
Considering runs with different field strengths and inclinations, $\varSigma$
varies in the same direction for every value of $\tauM$.

With increasing values of $\tauM$, the solutions become more unstable,
developing the typical stripes and clumps. Figure~\ref{fig:vartaum} shows this
for the standard case.  As an increase of $\tauM$ leads to a decrease of
density and gas pressure, this is consistent with the observation that the
amplitude of the instability increases at low $\beta$.

\subsection{Clumpy disks}
\label{sec:clumps}

\begin{figure*}[t]
\centering
\begin{tabular}{cc}
\parbox{.45\linewidth}{
\includegraphics[width=.9\linewidth]{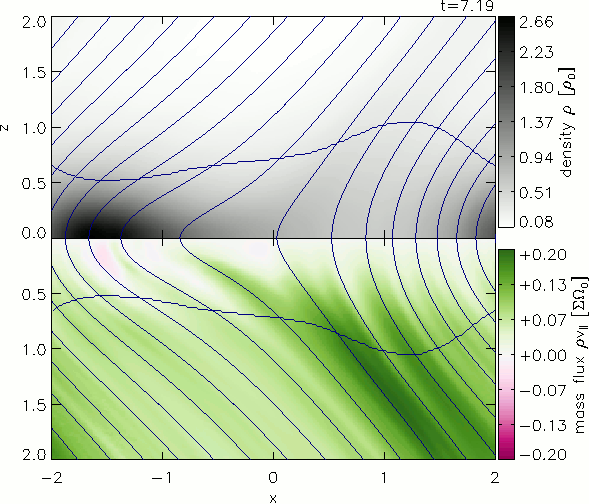} } &
\parbox{.45\linewidth}{
\includegraphics[width=.9\linewidth]{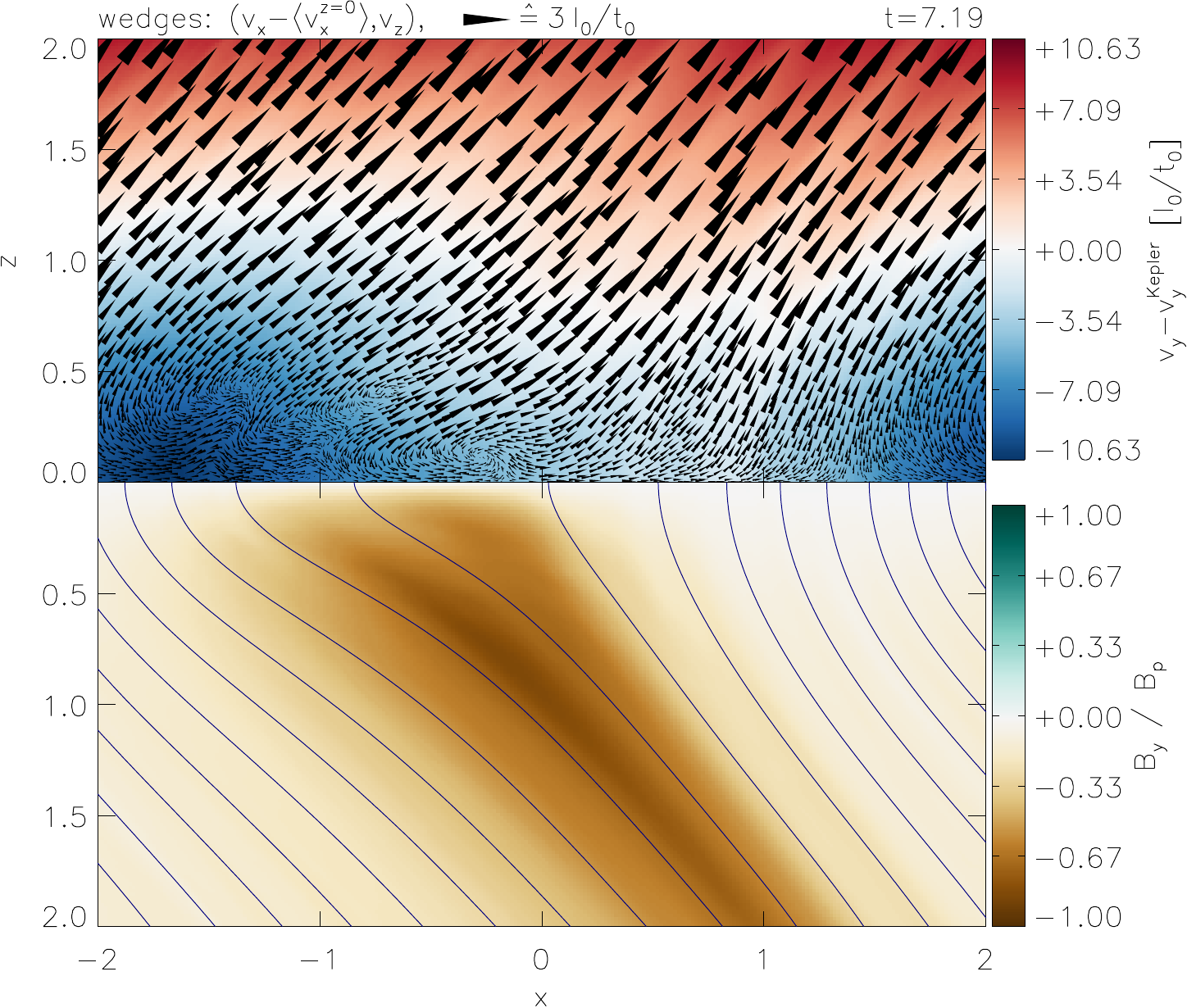} }
\end{tabular}
\caption{Snapshot of run $\beta1\iota50$ during the formation of the clump.
\textit{Left panel}: density (black/white) and field-parallel mass flux
(green/pink colors).  \textit{Right panel}: azimuthal velocity (sub-Keplerian
in blue and super-Keplerian in red) and azimuthal inclination of the magnetic
field (brown/turquoise colors).  The horizontal velocity, as measured in a
frame moving with the mean velocity at the midplane, is visualized with small
wedges whose areas are proportional to the velocity modulus.  Selected magnetic
field lines and the disk's scale height are overplotted in blue. See
Fig.~\ref{fig:Cimaps}c for a snapshot in the evolved state.}
\label{fig:C3maps}
\end{figure*}

\begin{figure}[t]
\centering
\includegraphics[width=.8\linewidth]{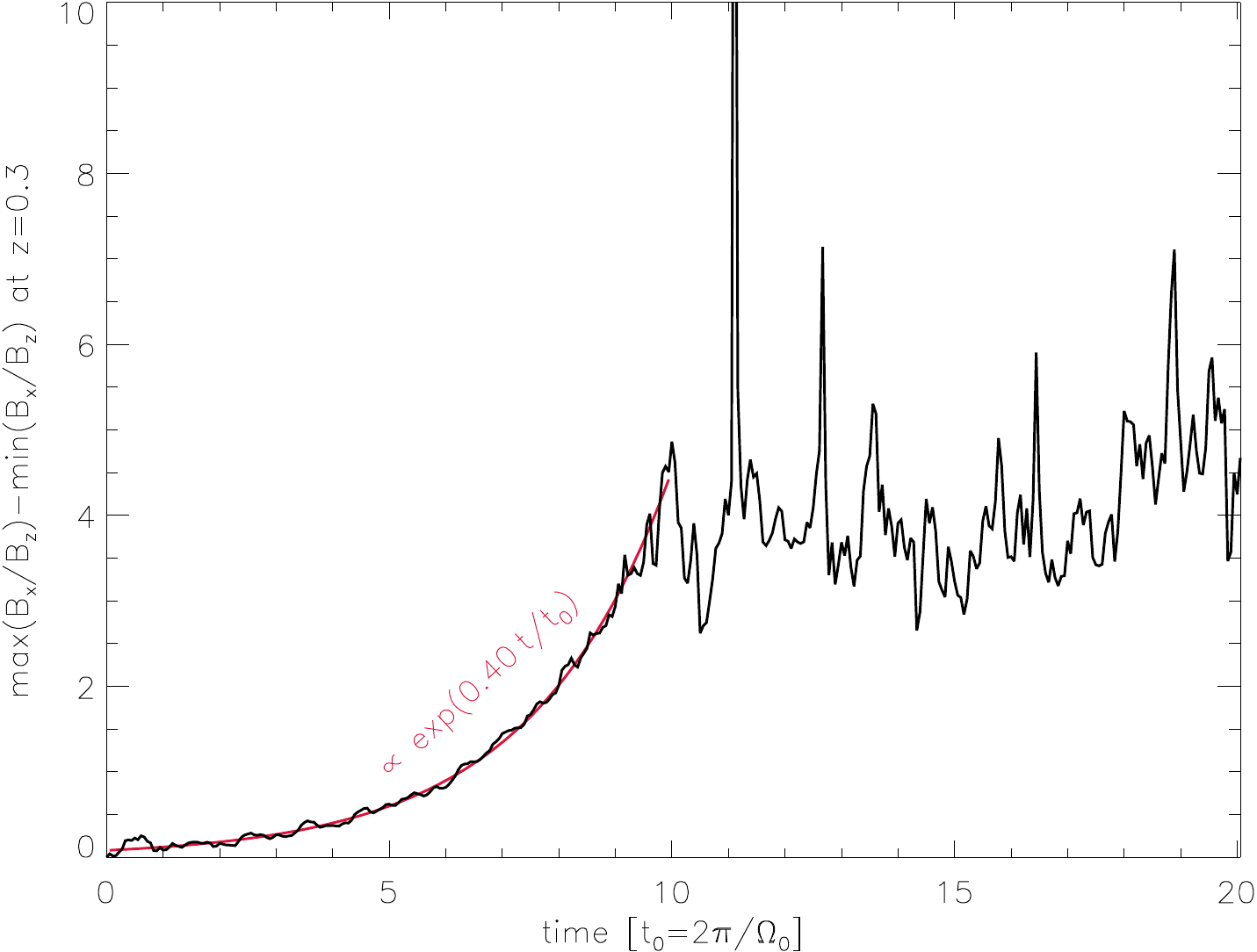}
\caption{Growth of the clump in run $\beta1\iota50$. As a measure for the
``clumpiness'', we use the range of poloidal field inclinations at a fixed
height in the disk.}
\label{fig:clgrow}
\end{figure}

\begin{figure}[t]
\centering
(a)\\
\includegraphics[width=.9\linewidth]{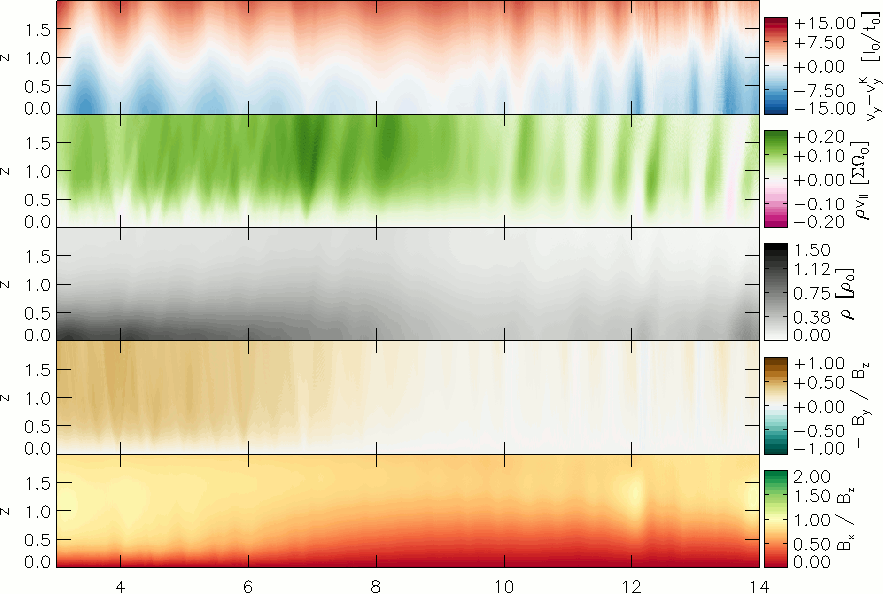} \\
(b)\\
\includegraphics[width=.9\linewidth]{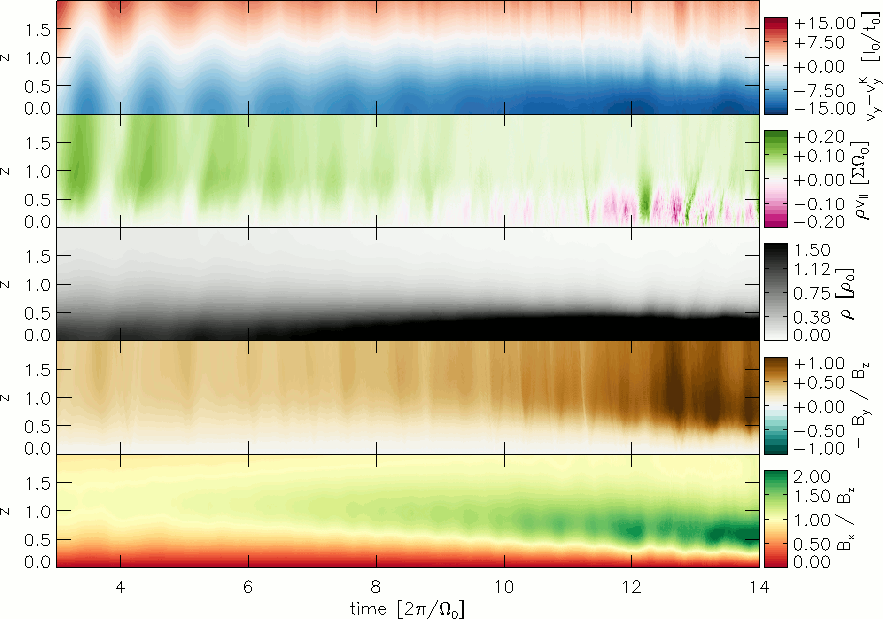}
\caption{Evolution of various quantities along two different field lines (a)
and (b) in run $\beta1\iota50$. From top to bottom: rotation velocity minus
Keplerian value, mass flux, density, field inclination in azimuthal direction
and field inclination in horizontal direction.}
\label{fig:C3fline}
\end{figure}

In some simulations, strong horizontal inhomogeneities in the density develop:
the disk condenses into one or several persistent ``clumps''.  A single
distinctive clump forms in run $\beta1\iota50$, see Fig.~\ref{fig:Cimaps}c.
During the formation, the outward mass flux is often higher along field lines
which are not anchored in the clump, see Fig.~\ref{fig:C3maps}.  The clump's
rotation is relatively slow (sub-Keplerian).  In its wake, plasma-$\beta$ is
high and the field at low heights is highly inclined towards the midplane.

We measure the growth of the clump by considering the inhomogeneity of the
inclination of the poloidal magnetic field inside the disk, at a height of
$z=0.3\approx0.4\mmean{H}$. The result is depicted in Fig.~\ref{fig:clgrow}.
The perturbation grows exponentially with a growth time of $2.5$ orbits.
Saturation is likely influenced by the horizontal periodicity of the
computational domain.  In simulation $\beta0.5\iota45$, which also develops a
single clump, we measure an exponential growth time of $4.2$ orbits by the same
method.

Fig.~\ref{fig:C3fline} shows the evolution of various quantities, (a) along a
field line for which the density at low heights decreases during clump formation
and (b) along a field line for which the density increases.  The difference in
density relates to an overall stronger mass flow (green colors) along field
line (a).  Line (b) develops a strong twist $B_y/B_z$ and a high inclination
towards the midplane at low heights.  Rightwards inclined stripes in the mass
flux plot indicate outward moving perturbations in the outflow. There are also
fan-shaped features which indicate perturbations that move down from the top
and are reflected at the midplane.

\subsection{Weak magnetic fields and MRI}
\label{sec:weakfields}

\begin{figure}[t]
\centering
\includegraphics[width=\linewidth]{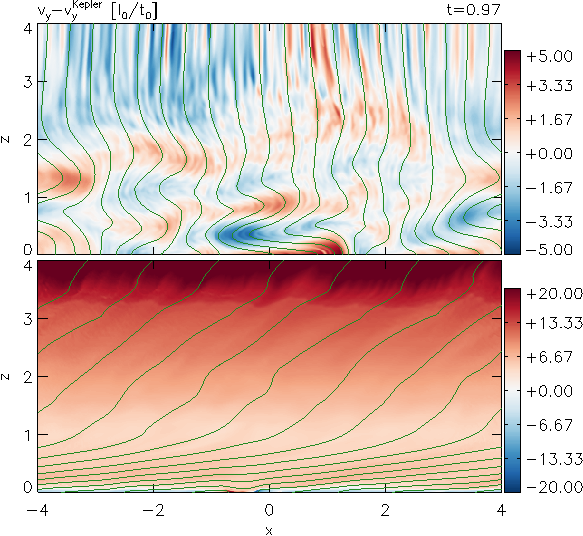}
\caption{Snapshots of the azimuthal velocity (sub-Keplerian in blue and
super-Keplerian in red colors) and magnetic field lines (green) in two
simulations with $\betain=100$: $\iota=90\degree$ in the \textit{top panel} and
$\iota=45\degree$ in the \textit{bottom panel}. The initial azimuthal velocity
was perturbed with white noise in both cases.}
\label{fig:C5b1e2n}
\end{figure}

\begin{figure}[t]
\centering
\includegraphics[width=\linewidth]{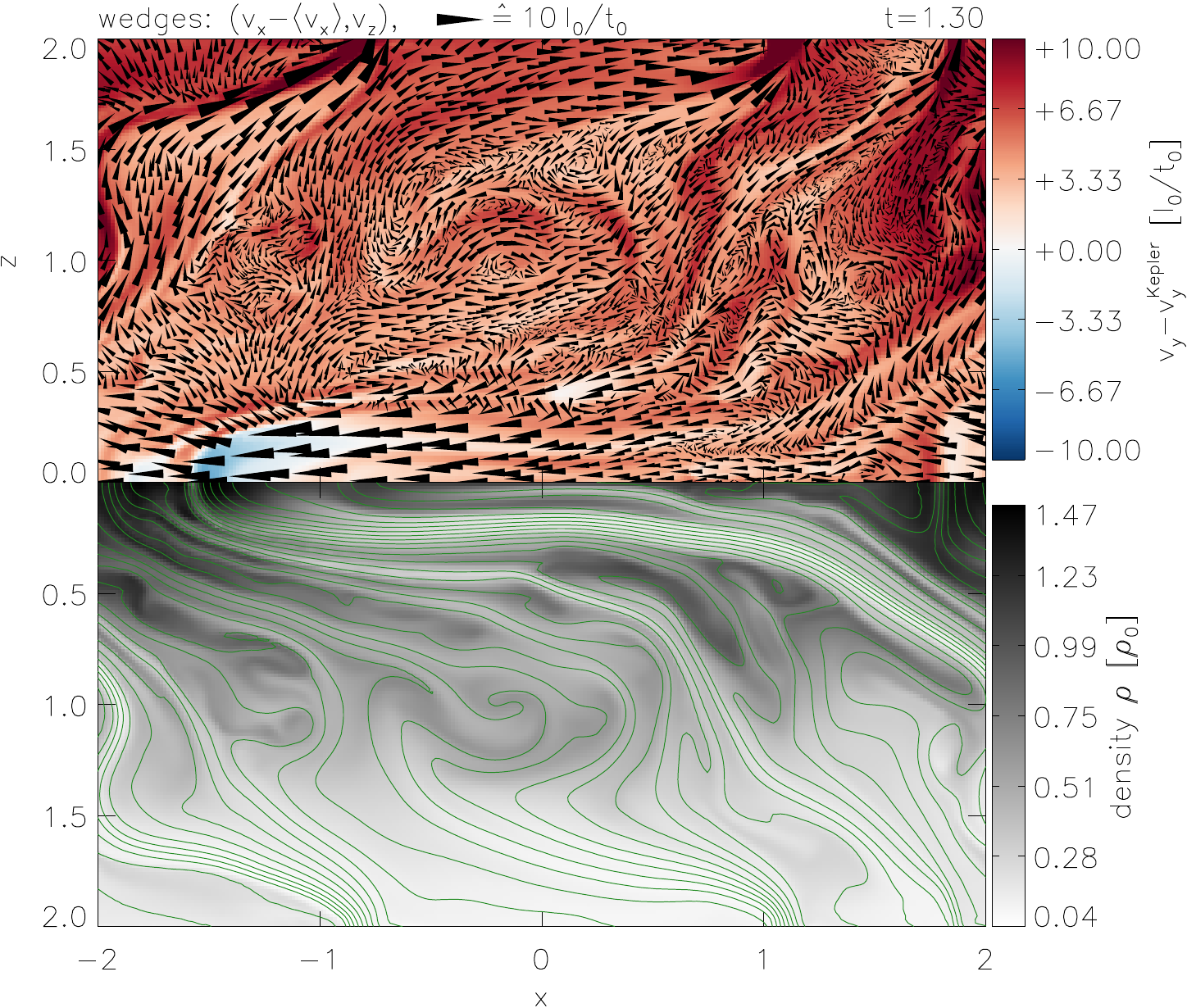}
\caption{Snapshots of velocity, density and magnetic field lines in a
simulation that begins in the evolved state of run $\beta1\iota45$
(Fig.~\ref{fig:C1map}) but with the magnetic field strength reduced by a
factor 10.  Wedges depict the poloidal velocity in a frame moving with the mean
horizontal velocity $\mean{v_x}=-3.8\,l_0/t_0$. Red and blue colors represent
super and sub-Keplerian azimuthal velocities, respectively.}
\label{fig:C5b1e2re}
\end{figure}

The upper magnetic boundary conditions, viz., a potential poloidal field, are
introduced under the assumption of strong magnetic fields with $\beta\ll 1$.
In the initial stratification, $\beta=1$ at $z=0$ in run $\beta1\iota45$ and
$\beta=1$ at $z=2.1$ (i.e., slightly above the upper boundary) in run
$\beta10\iota45$, which, of all cases presented above, is the simulation with
the weakest field.  The respective values for $\mmean{p}/\mmean{\pmag}$ in the
evolved state are $z=0.32$ and $z=0.44$; both are well below the upper
boundary.  That is, with the exception of perhaps an initial transient, the
boundary conditions are physically sound in the cases presented above.

The orderliness of the magnetic field is destroyed if the system is subject to
MRI. As expected, we see such cases in simulations with very low field
strengths.  Fig.~\ref{fig:C5b1e2n} shows snapshots of two simulations with
$\betain=100$ and no mass source ($\tauM\rightarrow\infty$). In both runs, the
initial azimuthal velocity $v_y$ was perturbed by random perturbations with an
amplitude of $1\,\cso\approx6.3\,l_0/t_0$.  In the first case (top panel),
$\iota=90\degree$. The initially vertical magnetic field is radially stretched
by axisymmetric MRI modes (compare, e.g., \citealt{2010Stone}).  The
perturbations grow strong within roughly an orbit.  In the second case (bottom
panel), $\iota=45\degree$.  The magnetic field becomes very flat at low heights
and the density develops a strong peak: $\mean{\rho(z=0)}/\mean{\rho(z=0.3)}
\approx 12$ at $t=0.97$. The field then reconnects at the midplane.

Runs with $\iota=45\degree$ and $\betain=12,100,10^{20}$ are all numerically
unstable (i.e., some values become too extreme for the numerical solver) and
terminate after $\simm 1/2$ orbit.  Common to all these cases is that the
magnetic field becomes very flat near the midplane.  Unlike in the low-$\beta$
cases, the reflecting condition does not make the field vertical near $z=0$.
For $\betain\rightarrow\infty$ (no magnetic field), as expected, the disk is
stationary and does not generate a wind.

To test whether the peculiar behavior at very weak magnetic field strengths is
tied to the initial state, we continued simulation $\beta1\iota45$ with the
magnetic field strength reduced by a factor 10. A snapshot of this run is shown
in Fig.~\ref{fig:C5b1e2re}. Due to the changed magnetic forces, the originally
stable system is out of equilibrium and oscillates. The magnetic field is
distorted by instabilities at all heights. After about 1 orbit, an elongated
magnetic structure develops near the midplane, moving inward fast.  The mean
plasma-$\beta$ is smaller than 1 for $z\gtrsim1.35$, i.e., the potential
condition at the upper boundary is still practical in this simulation.

\section{Summary and discussion}
\label{sec:discussion}

We have presented simulations of magnetocentrifugally accelerated winds in an
axisymmetric shearing box.  The box contains a thin disk with an ordered
poloidal magnetic field that is inclined away from the rotation axis.  Special
upper boundary conditions were used that allow a study of the coupling between
the disk and the angular momentum removing outflow.

For magnetic fields which are strong enough to suppress MRI, we find that
material is efficiently accelerated away from the disk.  To be able to follow
the disk evolution over many orbits, we replace lost material by a
time-independent mass source.  After several orbits, the wind enters a
quasi-steady equilibrium with the mass source. The strength of the wind, as
measured by how efficiently it limits the amount of material in the box in
spite of a constant resupply, increases with smaller magnetic field strengths
or smaller asymptotic inclinations of the poloidal magnetic field.  Angular
momentum is lost mainly through magnetic torques (Maxwell stresses). The disks
develop a radial inflow with sub-Keplerian rotation velocities near the
midplane.

In general, the outflows are not homogeneous and the magnetic field lines are
not uniform.  All cases develop instabilities in the mass flux on the shortest
length scales that are numerically resolved. In some cases, these simply
saturate. In other cases, the nonlinear development is much more dramatic,
leading to the formation of inward-moving density clumps.  Cases with high
relative field strengths (low plasma-$\beta$) develop especially strong clumps.
Such a clump grows exponentially with a growth time of a few orbits.  The
growth is tied to inhomogeneities in the strength of the outflow along
different magnetic field lines.  It is accompanied by a nose-shaped deformation
of the magnetic field, with the clump located at the tip of the nose.  Such a
massive instability has been proposed before in a simpler model of the
disk-flow connection \citep{2000Agapitou}.  Clumps may be a reason for the
variability seen in observations of accreting systems, for instance the hard
state noise in X-ray binaries \citep[e.g.,][]{2000Lin}.

At very low field strengths, the solutions become very different.
``Classical'' axisymmetric MRI modes develop if the field is vertical. The
modes grow strong within roughly an orbit.  With an inclined field, and despite
reflective symmetry, the field tends to become extremely flat near the
midplane.  This then leads to magnetic reconnection.

A major constraint of the simulations presented here is that 3D effects are
ignored. How do the 2D stripes in the wind turn into 3D structures?  We
circumvented the problem of how the material lost in the outflow is
replenished. In reality, this is presumably achieved through 3D interchange
processes. As suggested by 3D simulations of ``magnetically arrested'' flows
\citep{2003Narayan,2008Igumenshchev}, it is likely that high-amplitude clumps
also form in the 3D case. This would also provide justification for the
``clumpy field'' accretion of magnetic flux proposed by \citet{2005Spruit}.

\begin{acknowledgements}
The author thanks Henk Spruit for his time and support of this work, and Martin
Obergaulinger for providing his extremely versatile MHD code Aenus.  He thanks
Robert Cameron for his advice on Fourier transforms and kindly acknowledges
support through the Feodor Lynen Research Fellowship by the Alexander von
Humboldt Foundation.
\end{acknowledgements}

\bibliography{ref}

\end{document}